\documentclass[useAMS,usenatbib]{mn2e}

\usepackage{graphicx}
\DeclareMathAlphabet{\mathpzc}{OT1}{pzc}{m}{it}

\newcommand\mutotal{\mu_{\rm t}}

\title[]{Combining time delays and  image positions for quadruple lenses: a moment approach}
\author[Witt \& Mao] {Hans J. Witt$^{1,2}$ and Shude Mao$^{3,1,4}$ \thanks{E-mail: shude.mao@gmail.com}\\
$^{1}$ National Astronomical Observatories, Chinese Academy of
Sciences, Beijing, 100012, China \\
$^2$ Im Hollergrund 76, 28357 Bremen, Germany \\
$^3$ Physics Department and Tsinghua Center for Astrophysics, Tsinghua
  University, Beijing, 100084, China \\
$^{4}$ Jodrell Bank Centre for Astrophysics, University of Manchester, Manchester M13 9PL, UK
}
\begin{document}
\include{journaldefs}
\date{Accepted ...... Received ...... ; in original form......   }

\pagerange{\pageref{firstpage}--\pageref{lastpage}} \pubyear{2015}
\maketitle
\label{firstpage}

\begin{abstract}
Time delays in gravitational lenses can be used to determine the Hubble constant and
the lens potential. In future surveys, many gravitational lenses can be discovered, and their
time delays and image positions can in principle be measured. 
Using an elliptical power-law potential, we show that combinations of image positions
and time delays for quadruple lenses yield simple analytical expressions that are connected with observable quantities. These relations can be used to obtain the approximate axis ratio $q$, the Einstein radius and the slope. We apply this method to  RX J1131$-$1231, and show that our analytical results match the full numerical determinations approximately. Our approach can quickly determine rough values of lens parameters, which can then be used as initial guesses for further refinement through numerical modelling and may be useful for automated lens search in large surveys.
\end{abstract}
\maketitle


\section{Introduction}

Time delays in gravitational lenses provide an independent way of measuring the Hubble constant \citep{Ref64}.
Many studies have been performed in this direction. Observationally, two dozens or so lenses have their time delays accurately measured after painstaking efforts (e.g., \citealt{Tew13}). Theoretical models also become increasingly sophisticated. For example, a recent study by \citet{Suyu13} takes into account of the line of sight structures in addition to the lensing galaxy. Future surveys such as LSST will discover tens of thousands of lenses. In addition, hundreds of lensed supernovae will be found. A large fraction ($\sim 40\%$) will have crude time delays \citep{OM10}, while a smaller fraction of these ($\sim$ 400) will have well-measured values \citep{Liao15}. Such a large sample of lenses with time-delays can in principle be a very efficient way of constraining dark energy \citep{Lin11}.

The time delay for an isothermal potential or density distribution does not depend on the angular profile of the lensing galaxy \citep{WMK00}. Other than this, very few analytical results are known when the profile deviates from the isothermal shape. With 
thousands of time delays becoming available in the future, it is important to
understand better how the time delays depend on the density profile, and their
impact on the probability distribution of time delays. 
In this case one should not only focus on the time delays and the ratios
of time delays measured in double and quadruple lenses.
The ratio of time delays may depend on the lens model in a complex way. Therefore a diagram (of ratios) of time delays
may be difficult to interpret. Many Monte Carlo simulations
might be necessary  to understand their behaviour and the dependency of 
different lens models on the parameters. In this paper we show that it can be
useful to focus on the sum of time delays, and combinations
of time delays and image positions. These quantities are more
naturally connected to properties of the lensing galaxy such as the
axis ratio or the slope of the inner potential.

In this work we shall recall the properties of the elliptical power
law potential (\S\ref{sec:general}). While this lens potential is not as realistic as the elliptical density distribution (e.g., \citealt{ksb94, kks97}), it often resembles the latter when the ellipticity is small (\citealt{kk93}) and provides useful insights into the lensing properties. In  \S\ref{sec:moments}, we introduce a new moment approach through which we derive new connections between the time delays and their image positions. We apply the results to the
quadruple lens RX J1131$-$1231 in section \S\ref{sec:examples}, followed by a
short discussion in \S\ref{sec:discussion}. 
For ease of reading, we put most of the mathematical details in the appendices.

\section{Lensing Properties of an Elliptical power-law Potential}
\label{sec:general}

The elliptical power-law potential widely used in gravitational lensing
(e.g., \citealt{bk87, kb87}) is given by
\begin{equation}
\psi (x,y) = { a \over \beta} (x^2 + y^2 / q^2 )^{\beta / 2} 
\quad {\rm for } \quad 0 < \beta < 2
\label{eq:elliptical-potential}
\end{equation}
where $(x, y)$ are the (angular) lens position, $q$ is the axis ratio of the lensing galaxy ($0 < q \leq 1 $),
$\beta$ the slope of the potential and $a$ sets the angular scale of the lens. For $\beta =1$ we have the
elliptical isothermal sphere which has very simple properties (see below).

\subsection{Lens equation and Einstein radius}

The lens equation for the power-law potential is 
\begin{eqnarray}
\xi &=& x - {\partial \psi  \over \partial x } 
= x - a x (x^2+y^2/q^2)^{\beta/2-1}  \label{eq:xi}, \\
\eta &=& y - {\partial \psi \over \partial y} 
=  y - {a y \over  q^2} (x^2+y^2/q^2)^{\beta/2-1},
\label{eq:eta}
\end{eqnarray}
where $(\xi, \eta)$ are the source position. If the lens is circularly symmetric ($q=1$) and perfectly aligned with the source, the images form an Einstein ring with its angular scale given by
\begin{equation}
\theta_E=a^{1/(2-\beta)}.
\label{eq:thetaE}
\end{equation}

\subsection{Image Positions, Numbers and Magnifications}

The magnification of an image is generally given as
\begin{equation} \label{magnification}
\mu_i = {1 \over \det J} = \left. \left( {\partial \xi \over \partial x}
{\partial \eta \over \partial y} -
{\partial \xi \over \partial y} 
{\partial \eta \over \partial x} \right)^{-1} \right|_{x=x_i, y=y_i}
\end{equation}
where $(x_i, y_i)$ is the image position of the $i$th image in the
lens plane. The image positions where the magnifications are infinite form critical curves and their source positions form caustics. For an elliptical potential, the caustics are in general diamond-shaped. When a source falls inside the diamond caustics, we can have either four images for $0 < \beta \leq 1$ or five images for $1 < \beta < 2$. In the latter case, a faint, central image appears. 

The total magnification is given as the sum of all absolute magnifications
\begin{equation}
\mutotal = \sum_i | \mu_i |.
\end{equation}
For an elliptical power-law potential the sum of the signed magnification follows a simple relation
\begin{equation}
\sum_{i=1}^4 \mu_i \approx { 2 \over 2 -\beta }
\end{equation} 
for a quadruple system when the source is located inside the caustics. The result is exact for the cases $\beta = 1$ and $\beta = 4/3$ (\citealt{WM00} and \citealt{HE01}).

\subsection{Time Delay}

The time delay in gravitational lenses is given in general by
(e.g. \citealt{SEF92}) 
\begin{eqnarray} \label{eq:time_delay}
\Delta t &=& {D \over  2 c} (1+z_d) \tau(x,y),  \\
\tau(x,y) &\equiv& (x-\xi)^2 + (y-\eta)^2 - 2\psi(x,y), \label{eq:tau}
\end{eqnarray}
where $z_d$ is the redshift of the lens, and $D=D_{\rm d} D_{\rm s}/D_{\rm ds}$, 
with $D_{\rm d}, D_{\rm s}$ and $D_{\rm ds}$ are the
angular-diameter distances to the lens, to the source, and from the lens to
the source, respectively. 
Below we shall focus on $\tau$, and ignore all the cosmological dependences.

\section{Time Delays and Image Positions in Quadruple Lenses}
\label{sec:moments}

As mentioned before, when a source is located inside the caustics,
we may observe at least 4 images and possibly 3 time delays if the lensed source varies.
Such observations offer an exciting opportunity to probe the (inner) potential of the lensing galaxy.
In particular, future surveys will observe thousands of gravitational lensed systems
(see \citealt{OM10}). Some of theses systems might be fairly isolated
with negligible shear, which, as we show below, would be an ideal case to study
the potential slope and the axis ratio of the lensing galaxy in more details. In any case, the effects of a (small) shear enter linearly (see \S\ref{sec:appendixShear}), and so can be ignored as a first approximation.
 
For an isothermal potential with arbitrary angular profiles,  $\psi(r, \phi) = r F(\phi)$, 
the time delay between two images depend only on the
distances of the images $(x_i, y_i)$ from the galaxy centre, namely
\begin{equation} \label{simpleTD}
\Delta \tau_{i,j} = \tau(x_i, y_i) - \tau(x_j, y_j) =  r_j^2 - r_i^2
\end{equation}
with $r_i = \sqrt{x_i^2+y_i^2}$
(cf. \citealt{WMK00}).
In particular for an isothermal elliptical potential
it is rather straightforward to derive some invariants for the time delay.
We can extend the results of \citet{WM00} and obtain some simple
relations concerning the moments of the magnification and image positions.
Similar moment relations can be computed for the time delay as well. In particular
one should not consider only the ratio of the time delays, but also
the sum of all time delays of a system. We pursue this approach below; most of 
the mathematical details are collected in the Appendices.

\subsection{A Moment Approach}

We will develop here a moment approach using the time delays
to study the deviation from the ideal case of an elliptical isothermal sphere.
Let us  denote the moments of the time delay by
\begin{eqnarray}
{\cal T}_{x^m} &=& \sum_i \tau_i x_i^m,  \\ 
{\cal T}_{y^m} &=& \sum_i \tau_i y_i^m, 
\end{eqnarray} where
$\tau_i = \tau (x_i, y_i)$ is the time delay at the $i$-th image position.
Since the absolute time delays $\tau_i$ can not be measured directly
we relate them to measurable quantities by
\begin{eqnarray}
{\cal T}_{x^m} &=& \sum_i \Delta \tau_{i,1} x_i^m + \tau_1 \sum_i x_i^m  \\ 
{\cal T}_{y^m} &=& \sum_i \Delta \tau_{i,1} y_i^m + \tau_1 \sum_i y_i^m 
\end{eqnarray}
Now we use the $0$th-moment equation 
\begin{equation}
{\cal T}_0 = \sum_i \Delta \tau_{i,1} + n \tau_1
\end{equation}
to eliminate the unknown time delay of the first image $\tau_1$, where
$n$ is the total number of images. $n$ is usually 4 ($n=5$ if a central image is present).
Finally we can write
\begin{eqnarray}
\Delta {\cal T}_{x^m} &\equiv&
{\cal T}_{x^m} - {{\cal T}_0 \over n} \sum_i x_i^m \nonumber \\   
&=& \sum_i \Delta \tau_{i,1} x_i^m - {1 \over n} \sum_i \Delta \tau_{i,1} 
\sum_i x_i^m  
\end{eqnarray}
and
\begin{eqnarray}
\Delta {\cal T}_{y^m} &\equiv&
{\cal T}_{y^m} - {{\cal T}_0 \over n} \sum_i y_i^m   \nonumber \\
&=& \sum_i \Delta \tau_{i,1} y_i^m - {1 \over n} \sum_i \Delta \tau_{i,1} 
\sum_i y_i^m  
\end{eqnarray}
On the left side of the two previous equations we have now quantities
which can be expressed for example in terms of $\xi,\eta, q$ and $\beta$. On the right side, 
we have only terms involving the image positions
and the relative time delays, which can be measured directly for
a quadruple lens system.

\subsection{Analytical moment relations}

In the Appendix $\ref{sec:momentsTD}$ we present the detailed procedure through which
we find some analytic moment relations involving the relative time delay moments for a power-law potential. For simplicity, we normalise the coordinates by the Einstein radius, after which we have $a=1$. For the second order moments we find, when the source is located close to the origin with no shear ($\gamma = 0$), 
\begin{equation}
\sum_{i=1}^4 r_i^2 \approx 2 + {2 \over q^B } \quad {\rm with} \quad
B\equiv {2\beta \over 2-\beta}
\label{eq:iso}
\end{equation}
and
\begin{equation}
\Delta {\cal T}_{r^2} = \Delta {\cal T}_{x^2} + \Delta {\cal T}_{y^2} 
\approx - {2 \over B} \left( {1 \over q^B} - 1 \right)^2.
\label{eq:deltatr2}
\end{equation}
The approximations are valid as long as $-\Delta {\cal T}_{r^2} \gg 10 (\xi^2 +
\eta^2)$. 
Alternatively we obtain for the case of an isothermal sphere
with an on-axis shear ($\beta = 1, B = 2$)
\begin{equation}
\sum_{i=1}^4 r_i^2 \approx {2 \over (1+\gamma)^2} + {2 \over (1-\gamma)^2 q^2 } 
\end{equation}
and
\begin{equation}
\Delta {\cal T}_{r^2}  
\approx 
 {-1 \over (1+\gamma)^3} 
\left( 1  - {(1+\gamma) \over (1-\gamma) q^2} \right) 
\left( 1  - {(1+\gamma)^2 \over (1-\gamma)^2 q^2} \right).
\label{eq:deltaTr2}
\end{equation}
Notice that the previous four equations are rotationally invariant, and thus do not
depend on whether we know the position angle of the lens galaxy.

For fairly round system with $q \rightarrow 1 $ the image positions
are very close to the Einstein ring so that the solutions approach 
$r_i \rightarrow 1$
and the sum  of the squared image positions $\sum_i r_i^2 \rightarrow 4$.
The time delay between the images becomes smaller and smaller so that
the relative time delay moment vanishes ($\Delta {\cal T}_{r^2} \rightarrow 0$). 
For more elliptical systems
the sum of the squared image distances to the galaxy centre increases
with smaller $q$. The more elliptical the system becomes
the more the image positions are located further away from the Einstein ring.
In particular in clusters where one expects $1 < \beta \la 1.5$
this effect is considerably amplified since for $\beta = 4/3$,
$B=4$ and for $\beta = 3/2$, $B=6$. Therefore in clusters the source location
plays less of a role and the system is mostly determined by $q$ and $\beta$.
Similar effects can be observed if we have an isothermal sphere
($\beta =1$) plus shear. The shear acts like an amplifier for smaller $q$.
Therefore the shear $\gamma$ may emulate a more elliptical system than the ellipticity of the galaxy actually indicates.

For the relative time delay moment $\Delta {\cal T}_{r^2}$
the term becomes small when the system is nearly round and $q\la 1$.
In this case the time delay is dominated by the source position and the
approximation is fairly bad. For these cases we need to take
the source positions into account. This would mean we need
to solve a nonlinear set of equations and take the first
order moments into account as well. Also in the case when the source
is located near a cusp $\xi$ or $\eta$ might not be negligible any more.
However, for rather elliptical system with $q \la 0.75$ and $\beta \approx 1$
the source position
plays less a role, so that the approximations hold fairly well
in the case of galaxy lensing.
Such a system would be suitable to determining the axis ratio $q$ and
the slope of the potential $\beta$.

Based on numerical simulations, the two sets of equations (eqs. \ref{eq:iso} and \ref{eq:deltatr2}) seem to be rather robust and
they depend mainly on the uncertainty of $D$ in eq.  (\ref{eq:time_delay})
which can be determined if the distances
to the lensing galaxy and the source, i.e.
the redshifts $z_d$, $z_s$ and the Hubble constant $H_0$, are known.
We note here that it is also possible to derive similar equations
if we use the ratios of the time delay moments or the image positions 
in the form of
\begin{equation}
{\sum_i x_i^2 \over \sum_i y_i^2} \approx {(1-\gamma)^2 \over (1+\gamma)^2} q^B 
\approx- {\Delta {\cal T}_{x^2} \over \Delta {\cal T}_{y^2} } 
\end{equation}
(see Appendix A). The equation is valid for the case $\gamma = 0$ or
$B=2, \beta=1$.
Such equations have the advantage they are ratios and thus scale-free. However, unlike eqs. (\ref{eq:iso}) to (\ref{eq:deltaTr2}), to apply this relation, we need
to transform the observed system into an unrotated coordinate system aligned with the symmetry axes of the lens.
The reader should consult the Appendix \ref{sec:rotation} for the procedure to do this.

Furthermore, when the external shear is negligible for an observed quadruple lens
we can derive an approximate estimate on the slope $\beta$ of the potential
using the four image positions and the three time delays. The details can be found in the Appendix
\ref{sec:slope}.

\section{Application to the lens system RX\,J1131$-$1231}

\label{sec:examples}

\begin{table}
        \centering
        \caption{Image, galaxy positions and time delays for RX J1131$-$1231. The time delay $\tau_{AX}$ is the time delay between image $A$ and $X=B, C, D$. G denotes the lens galaxy.}
        \label{tab:RXJ}
        \begin{tabular}{cccc}
                \hline
        Object & X $[^{\prime\prime}]$ & Y $[^{\prime\prime}]$ & $\tau_{AX}$ [days] \\
                \hline
                A & $+0.000\pm 0.000$ & $+0.000\pm 0.000$ & 0 \\
                B & $+0.032\pm 0.002$ & $+1.188 \pm 0.002$ & $0.7 \pm 1.0$ \\                
                C & $-0.590\pm 0.003$ & $-1.120\pm 0.003$ & $0.0 \pm 1.3$ \\
                D & $-3.112\pm 0.003$ & $+0.884\pm 0.003$ & $93.8 \pm 1.5$ \\
                G & $-2.016\pm 0.002$ & $+0.610\pm 0.002$ & $- $\\
                \hline
        \end{tabular}
\end{table}

\begin{figure}
\includegraphics[scale=0.40,bb=0 0 385 567]{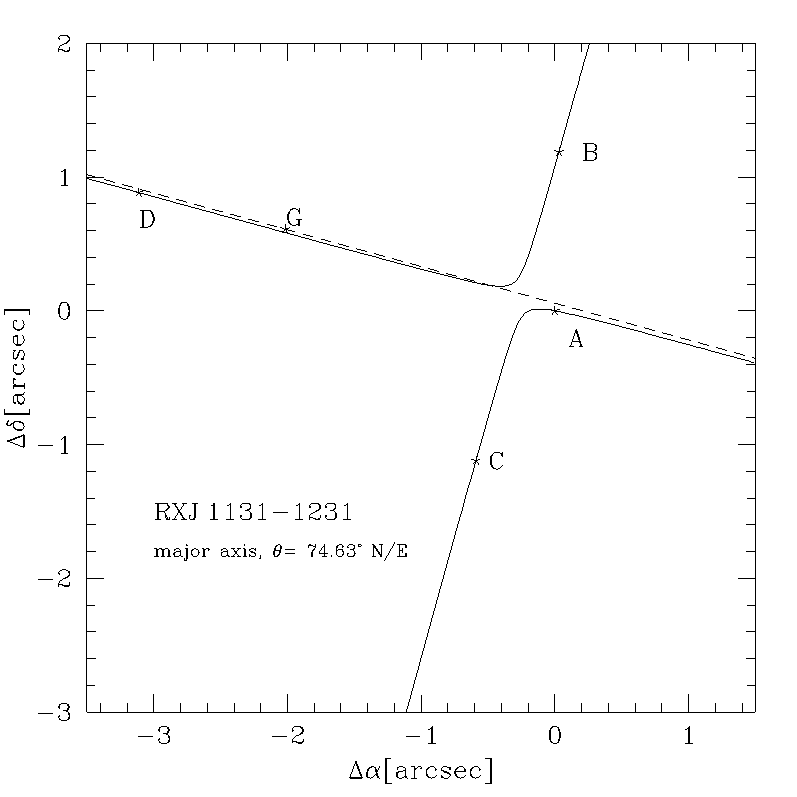}
\caption{
Image positions and galaxy position of RX J1131$-$1231.
The hyperbola line indicates the confinement of the galaxy and source
positions if a pure elliptical potential (plus an on-axis shear) is assumed.
The minimum distance of the galaxy position to the solid line
is $\Delta_{\rm min} = 0.025^{\prime\prime}$ indicating that the lens system can
not be perfectly fitted by any elliptical potential.
The major axis (dashed line) is rotated by $\theta_{N/E} = -74.63^\circ$ which
corresponds to a rotation angle of $\theta = -15.37^\circ$ from the $x$-axis
(cf. Appendix \ref{sec:rotation}). 
}
\label{fig:RXJ1131}
\end{figure}

In this section, we apply our formalism to the quadruple lens RX J1131$-$1231.  
The lensing galaxy is an elliptical at redshift 0.295 and the source redshift is
0.658. 
The large separation ($\approx 3$ arcsec) makes study of the system relatively 
simple (see \citealt{Cla06}). 
The image, galaxy positions and the measured time delays (\citealt{Tew13})
are listed in Table \ref{tab:RXJ} and Figure \ref{fig:RXJ1131}.
For the cosmology we use a matter density of $\Omega_{\rm m}= 0.3$, a cosmological
constant of $\Omega_\Lambda = 0.7$, and $H_0=70{\rm km\, s^{-1}\, Mpc^{-1}}$.

We first check whether the power-law potential gives a reasonable fit to the data. \cite{Wi96} showed that in such case, the images and the lensing galaxy are located on the same hyperbola-like curve. Furthermore, the position angle of the lens galaxy can also be inferred; the procedure to do this is outlined in the Appendix \ref{sec:rotation}.

The results are shown in Fig. 
\ref{fig:RXJ1131}. Indeed the galaxy and images fall approximately on the curve, with the maximum distance from the hyperbola being about 0.025 arcsec.  This should be compared with the positional accuracy of about a few mas (see Table \ref{tab:RXJ}). Our simple model appears to provide a reasonable approximation. Note that, from this model we can also derive the position angle of the lens galaxy, and obtain good agreement with the singular isothermal ellipsoid model in \citet{Cla06}.

Recall that we implicitly assumed the angles are 
normalised by the Einstein radius, $\theta_E$ (cf. eq. \ref{eq:thetaE}), i.e., $a=1$.
After normalisation, for $\beta=1, B=2$, we find for the sum of the squared image positions
\begin{equation}
\sum_{i=1}^4 r_i^2 = { 15.266 (1^{\prime\prime})^2 \over \theta_E^2 } \approx
2 + { 2 \over q^2} 
\end{equation} 
For the moment of the time delays in eq. (\ref{eq:deltatr2}) we need 
to relate the normalised units to the observed time delays and image positions.
We find that  $D/(2c) (1+z_d) = 32.5$ days$/ (1^{\prime\prime})^2$, and so
\begin{equation}
\Delta {\cal T}_{r^2} = { -237.8 \, {\rm days} \, (1^{\prime\prime})^4 \over 32.5
 \, {\rm days} \quad \theta_E^4 } \approx  - \left( { 1\over q^2} -1 \right)^2 
\end{equation}
From these two equations  we find $q=0.69$  and 
$\theta_E = 1.57^{\prime\prime}$.
These values are roughly consistent with the results from the singular isothermal ellipsoid 
model in Table 4 in \citet{Cla06}, for which
they gave $\theta_E=1.82\arcsec$, and $\epsilon=0.45$. The latter corresponds to an axis ratio $q=0.61$
by the relationship
\begin{equation} \label{eq:epsilon}
q=\sqrt{1-\epsilon \over 1 +\epsilon} \, .
\end{equation}
Our value differs from theirs only by approx. $10\%$. The differences are mostly due to the
neglected source positions in eqs. (\ref{eq:iso}) and (\ref{eq:deltatr2})
and the differences between elliptical density 
and elliptical potential models. An additional error of a few percent enters 
from the uncertainty in the measured time delays.

\subsection{The impact of the slope of the potential on the model}

Such a simple elliptical isothermal model as discussed above
can account only roughly the observed time delays. However,
the model above would predict roughly $\Delta t_{A,B} = -3$ days,
$\Delta t_{A,C} = -19$ days and $\Delta t_{A,D} = 103$ days
using eq. (\ref{simpleTD}) directly. This is certainly quite far away 
($\sim 10 \sigma$) from the observed time delays as listed in Table \ref{tab:RXJ}. It even predicts that
the order of arrival time would be C, B, A, D.

In this section we will adjust the slope of the potential
$\beta$ in order to better account for the observed time delays, and see
how the change affects the predicted axis ratio $q$. In the Appendix \ref{sec:slope} we show how the slope 
of the potential can be inferred just from the observed image positions
and time delays for a quadruple lens when the external shear is 
negligible. For the case of RX J1131$-$1231, we obtain $B=2.65$ which corresponds
to $\beta=1.14$.

Using eqs. (\ref{eq:iso}) and (\ref{eq:deltatr2}) 
we have
\begin{equation}
{ 1 + q^B \over 1 - q^B} 
\approx { C \over \sqrt{B} } \quad {\rm with} \quad 
C \equiv {1 \over \sqrt{-2 \Delta {\cal T}_{r^2}} }
\sum_{i=1}^n r_i^2 = 3.99
\end{equation}
which depends entirely on the observed image positions and time delays
of a quadruple lens. Finally we can deduce the axis ratio by
\begin{equation}
q \approx \left( { C - \sqrt{B} \over C + \sqrt{B} } \right)^{1 / B} = 0.72
\end{equation}
This value differs by $5\%$ of the previous derived value ($q \approx 0.69$),
and thus rather robust. For the Einstein radius we deduce 
$\theta_E = 1.61^{\prime\prime}$ which differs about $3\%$ from the previous model
($\theta_E=1.57^{\prime\prime}$).
In comparison, this model with free slope predicts
$\Delta t_{A,B} \approx -1.1$ days,
$\Delta t_{A,C} \approx -1.2$ days and $\Delta t_{A,D} \approx 93$ days, 
closer to the
observed values. Note that $\Delta t_{A,B}$ and $\Delta t_{A,C}$ depend
rather sensitive on the goodness of the fit.
Even this model requires further refinement, for example, 
the inclusion of an external
shear will better fit the observed image positions.
However, this may even change our derived axis ratio $q$ by a few percent.

\subsection{Relations to the Hubble constant}

Finally we would like to point out a simple relation between the
ellipticity $\epsilon$ of the lensing galaxy and the Hubble constant $H_0$.
Using eqs. (\ref{eq:iso}), (\ref{eq:deltatr2}) and (\ref{eq:epsilon})
we can write
\begin{eqnarray}
{ \Delta {\cal T}_{r^2} \over \left( \sum_i r_i^2 \right)^2 } &=&
{ 2 c \, \Delta T_{R^2}  \over D (1+z_d) \left( \sum_i R_i^2 \right)^2 } 
\nonumber \\
& = &{ \Delta T_{R^2}  \over \left( \sum_i R_i^2 \right)^2 }  H_0 f(z_d,z_s) 
\approx   
- {\epsilon^2 \over 4 }
\end{eqnarray}
where $R_i$ and $\Delta T_{R^2}$ denote the observed (unnormalised)
quantities and 
\begin{equation}
f(z_d, z_s)=\frac{2 c/H_0}{D(1+z_d)}
\end{equation}
is a function which just depends
on the redshifts of the source and the lensing galaxy, the matter density and the cosmological constant, but not on the Hubble constant. However, this relation holds only if the lensing galaxy is not distorted by shear.

\section{Summary and discussion}
\label{sec:discussion}

In the future, wide-field surveys such as the LSST will discover tens of thousands of new gravitational lenses (\citealt{OM10}), and for hundreds of these, their time delays will be measured accurately \citep{Liao15}. Such systems would be useful for independent determinations of the Hubble constant and understanding the evolution of dark energy. 

In this paper, we have developed a new approach by studying the cross-moments of the time delays and image positions with
an elliptical power-law potential. We find that combinations of image positions
and time delays for quadruple lenses yield simple analytical expressions that are connected with observable quantities, which can be used to obtain the approximate axis ratio $q$, the Einstein radius and the slope $
beta$. We apply this method to  RX J1131$-$1231, and show that our analytical results match the full numerical determinations approximately. 

These results may be useful for the following applications. First, the derived axis ratio and slope 
can be used as initial guesses for further refinement through numerical modelling of the lenses. Second, the current Time Delay Challenge One (TDC1) \citep{Liao15} uses only the light curves, but not the image positions. In the next challenge, it is likely that the image positions will be taken into account in determining the time delay. Our moment approach may be useful in rejecting ``catastrophic outliers" in the time delay. Third, our simple results may be useful for automated discovery of lenses in large surveys (\citealt{Mars09}; \citealt{Chan14}). For example, the fact that the lens positions fall on an approximate parabola may be useful in rejecting unrealistic image configurations without doing much modelling.

Our approach is clearly not accurate enough for each individual system. However, if it is unbiased then it may still be useful for inferring the cosmological parameters (such as the Hubble constant $H_0$) statistically, similar to the approach taken by \citep{Ogur07}. It will be interesting to study this issue using a mock sample of lenses in a future work.

\section*{Acknowledgement}
This work has also been supported by the Strategic Priority Research
Program The Emergence of Cosmological Structures of the Chinese Academy of Sciences
Grant No. XDB09000000, and by the National Natural Science Foundation of China (NSFC)
under grant number 11333003 (SM).

\bibliographystyle{mn2e}
\bibliography{./lens}

\appendix

\section{Moments of the time delay}
\label{sec:momentsTD} 

\subsection{The elliptical power-law potential}

For the elliptical potential given in eq. (\ref{eq:elliptical-potential}), the normalised time delay map is given by
\begin{equation}
\tau(x,y) = (x-\xi)^2 + (y-\eta)^2 - {2 a\over \beta} (x^2 +y^2/q^2)^{\beta / 2},
\end{equation}
for $0<\beta<2$. We show below that the time delay moments for $\beta = 1$
and $\beta=4/3$ can be  derived exactly. These
results are then used to extrapolate to other values of $\beta$ 
with $0 < \beta < 2$. Note that we use $a=1$ 
in this Appendix for convenience.

\subsubsection{Preliminaries}

Let us denote $\tau_i = \tau(x_i,y_i)$ where
$(x_i,y_i)$ are the image positions of the $n$ images  $i=1,...,n$ if the
the source is located inside the diamond caustic. 
Note that we have four images ($n=4$) for $0 < \beta \leq 1$, while an additional central fifth image
appears ($n=5$) when $1 < \beta < 2$.

For the case $\beta=1$ and $\beta=4/3$ we can readily separate the 
solutions of the lens equation $(x,y)$ into two polynomials. 
\begin{equation}
g_x(x,\xi,\eta) = \sum_{i=0}^n a_i x^i \quad,\quad 
g_y(y,\xi,\eta) = \sum_{i=0}^n b_i y^i \quad,\quad 
\end{equation}
with $n=4$ for $\beta=1$ and $n=5$ for $\beta=4/3$
(cf. \citealt{WM00} and \citealt{HE01}).
The moments of the image positions can now be related
to the coefficients of the polynomials:
\begin{eqnarray}
S_x &\equiv& \sum_{i=1}^n x_i = -{a_{n-1} \over a_n}, \\
S_{x^2} &\equiv& \sum_{i=1}^n x_i^2 = \left( {a_{n-1} \over a_n} \right)^2 - 2
{a_{n-2} \over a_n }, \\
S_{x^3} &\equiv& \sum_{i=1}^n x_i^3 = - \left( {a_{n-1} \over a_n} \right)^3
+ 3 { a_{n-2} a_{n-1} \over a_n^2 } - 3 { a_{n-3} \over a_n},\\
S_{x^4} &\equiv& \sum_{i=1}^n x_i^4 = \left( {a_{n-1} \over a_n} \right)^4
-4 {a_{n-2} a_{n-1}^2 \over a_n^3} \nonumber \\ & & 
+ 4 {a_{n-3} a_{n-1} \over a_n^2 }
+2{a_{n-2}^2 \over a_n^2} -4 {a_{n-4} \over a_n }.
\end{eqnarray}
For the $y$ component, analogous results are obtained: we just need to replace the $a_i$ by $b_i$.
For mixed terms we can use the relationship of the image positions
\begin{equation}
(\xi - x_i) y_i = q^2 x_i (\eta - y_i)
\end{equation}
which is valid for any elliptical potential (cf. \citealt{Wi96}).
Now we can write
\begin{equation}
S_{xy} \equiv \sum_{i=1}^n x_i y_i = {\xi \over (1-q^2) } S_y 
- {q^2 \eta \over (1-q^2) } S_x.
\end{equation}
Using these relations we are able to compute any desired
mixed terms like $S_{xy^2}$ and so on successively.

Recalling eq. (24) in \citet{WMK00} , we can write for the time delay
\begin{equation}
\tau (x,y) = \xi^2 +\eta^2 + { (2-B) \over B } (\xi x +\eta y)
- {2 \over B} (x^2+y^2),
\end{equation}
where we expressed $\beta \equiv 2B /(2+B)$ in terms of $B$.
This equation is in particular valid because $\psi$ obeys the
partial differential equation $\beta \psi = x \psi_x + y \psi_y$.
This enables us to express the potential $\psi$ in terms of the lens equation.

For the moments of the time delay, we can write now
\begin{equation}
\sum_{i=1}^n \tau_i = n (\xi^2+\eta^2) + { (2-B) \over B}  (\xi S_x+ \eta S_y) 
- {2\over B} (S_{x^2} + S_{y^2})
\end{equation}
and
similarly, higher order moments can be written as
\begin{eqnarray}
\sum_{i=1}^n \tau_i x_i &=& S_x (\xi^2+\eta^2) + { (2-B) \over B}  (\xi S_{x^2} + 
\eta S_{xy})  \nonumber \\
& & -{2\over B} (S_{x^3} + S_{xy^2})
\end{eqnarray}
and so on.

\subsubsection{Moment relations for elliptical power-law potentials}

In this section we present the results for the moments
for the power-law potential. These results are only
exact for the case $\beta=1$ and $\beta = 4/3$. However, for different
$\beta$ values ($0 < \beta < 2$) numerical results differ only by a few
percent or less if $\xi$ and $\eta$ are small (the source is sufficiently well aligned with the line of sight).

For convenience we define the following quantities
\begin{equation}
B \equiv {2 \beta \over 2 -\beta} \quad {\rm and } \quad 
 Q \equiv \sqrt{1 -q^2} .
\end{equation}
Note that $B=2$ for $\beta =1$ and $B=4$ for $\beta=4/3$.

For the sum of the time delay we obtain  in general
\begin{equation}
{\cal T}_0 = \sum_{i=1}^n \tau_i = - {4 \over B} (1 + {1 \over q^B})
- {2 q^2 \over Q^2} \xi^2 + {2 \over Q^2 } \eta^2
\end{equation}
and for the first order moments we end up with
\begin{eqnarray}
{\cal T}_{x}&=& -\xi \left[ 5 + {2\over B} + { 4 \over B Q^2 q^B }
\right], \nonumber \\ & &
- 2\xi^3 {q^2 \over Q^4 } + 2 \xi \eta^2 {(1+2 q^2 ) \over Q^{4}}   
\\
{\cal T}_{y}&=& - \eta \left[ { 5 \over q^B } + {2 \over B q^B} 
+ {4 q^2 \over B Q^{2} } \right] \nonumber \\ &&
+ 2 \xi^2 \eta { q^2 (2+q^2) \over Q^{4} } -2 \eta^3 {q^2 \over Q^{4}}. 
\end{eqnarray}
For the second order moments we obtain
\begin{eqnarray}
{\cal T}_{x^2} &=&  -{4 \over B} -[4+ {2 \over B}+ { 5 B \over 2}] \xi^2 
- {4 \xi^2 \over B q^B Q^{4}} \nonumber \\ 
& & + [{4 q^2 \over B Q^{4}} + {2\over Q^{2}}] \eta^2 \nonumber \\
& & - {2 q^2 (\xi^2-\eta^2)^2 \over Q^{6} } 
    + 2 \xi^2 \eta^2 {(1+q^2)^2 \over Q^{6} }, \\
{\cal T}_{y^2} &=& - {4 \over B q^{2B}} + {2 q^2 \over (1-q^2)^2 } \xi^2
\nonumber \\
& & -[4+ {2 \over B}+ { 5 B \over 2}] {\eta^2 \over q^B }
- {4 q^4 \eta^2 \over B Q^{4} } \nonumber \\
& & + {2 q^4 (\xi^2-\eta^2)^2 \over Q^{6} } 
    - 2 q^2 \xi^2 \eta^2 {(1+q^2)^2 \over Q^{6} }.
\end{eqnarray}
Now taking the results for the moments of the image
positions (which are only exact for the cases $\beta = 0, 1$ and $4 / 3$) 
\begin{eqnarray}
S_x &=& {\xi \over Q^{2}}  ((3+ {B \over 2})-(1+ {B \over 2})q^2), \\
S_y &=& {\eta \over Q^{2}} ((1+ {B\over 2})-(3+ {B\over 2})q^2),\\
S_{x^2} &=& 2  + (1+{B \over 2}) \xi^2 + { 2\over Q^{4} } \xi^2 
           -{2 q^2 \over Q^{4}} \eta^2, \\
S_{y^2} &=& { 2  \over q^B} + (1+{B \over 2}) \eta^2
         - { 2 q^2 \over Q^{4} } \xi^2+{2 q^4 \over Q^{4}} \eta^2,
\end{eqnarray}
we obtain the relative moments for the time delay for quadrupole lenses
\begin{eqnarray}
\Delta {\cal T}_{x} &=&  {\cal T}_{x} - {\cal T}_0 {S_x \over 4}, \\
\Delta {\cal T}_{y} &=&  {\cal T}_{y} - {\cal T}_0 {S_y \over 4}, \\
\Delta {\cal T}_{x^2} &=&  {\cal T}_{x^2} - {\cal T}_0 {S_{x^2} \over 4} 
\approx - { 2 \over B } + { 2 \over B q^B }, \\
\Delta {\cal T}_{y^2} &=& {\cal T}_{y^2} - {\cal T}_0 {S_{y^2} \over 4} 
\approx - { 2 \over B q^{2B} } + { 2 \over B q^B }.
\end{eqnarray}
The latter two equations hold only if $\xi,\eta \ll 1$ are small.
We did not expand the above expressions further because they follow
a rather complicated pattern from which one does not gain much insight.
However, it is interesting to note that
\begin{equation}
\Delta {\cal T}_{r^2} = \Delta {\cal T}_{x^2} + \Delta {\cal T}_{y^2} 
\approx - {2 \over B} \left( 1 - {1 \over q^B} \right)^2
\end{equation}
and
\begin{eqnarray}
\sum_{i=1}^4 r_i^2 &=& S_{x^2} + S_{y^2} = 2 + { 2 \over q^B} 
+ (1+{B \over 2}) (\xi^2+\eta^2) \nonumber \\ \label{sumri2}
& & + { 2 \over Q^{2} } \xi^2 
- {2 q^2 \over Q^{2} } \eta^2 \approx 2 + { 2 \over q^B} 
\end{eqnarray}
for small $\xi$ and $\eta$. The expressions depend only
on the distances of the image positions to the galaxy centre
and the relative time delays between the images, and so are rotationally invariant.
In addition the ratios of the relative time delay moments satisfy
\begin{equation}
- {\Delta {\cal T}_{x^2} \over \Delta {\cal T}_{y^2} } \approx q^B \approx
{S_{x^2} \over S_{y^2} }.
\end{equation}
Such equations can be used to cross check our results. However in this case we
need to derive the coordinates of the unrotated system (see Appendix \ref{sec:rotation}).

\subsubsection{Moment relations for an isothermal sphere plus an on-axis shear}
\label{sec:appendixShear}

For the case of an isothermal sphere plus an on axis shear (\citealt{WM97})
we can derive the moments in a similar way as in Appendix \ref{sec:momentsTD}
or by using the scaling properties of the lens equation in Appendix
\ref{sec:OnAxisShear}.
We just state here the approximate results for convenience.
For the image position moments we have
\begin{equation}
S_{x^2} \approx { 2\over (1+\gamma)^2 } \quad {\rm and} \quad
S_{y^2} \approx { 2\over (1-\gamma)^2 q^2 } 
\end{equation} 
and for the time delay moments we obtain
\begin{equation}
\Delta {\cal T}_{x^2} \approx {-1 \over (1+\gamma)^3 }+ 
{1 \over (1-\gamma)(1+\gamma)^2 q^2 },
\end{equation}
and
\begin{equation}
\Delta {\cal T}_{y^2} \approx {-1 \over (1-\gamma)^3 q^4 }+ 
{1 \over (1-\gamma)^2(1+\gamma) q^2 }.
\end{equation}
We note here that we only obtain rather simple results
for the relative time delay moments 
for the case of an isothermal sphere when $\beta =1$ and $\gamma=0$.
For completeness the results are as follows
\begin{eqnarray}
\Delta {\cal T}_{x} &=& -5\xi -  {q^2 \over Q^{2} }\xi^3 
- {5 \over Q^{2}} \xi\eta^2, \\
\Delta {\cal T}_{y} &=& -{5 \over q^2}\eta 
-{ 5 \over Q^{2} } \xi^2\eta  - {1 \over q^2 Q^{2} }\eta^3 \\
\Delta {\cal T}_{x^2} &=& -1 + {1 \over q^2} -10\xi^2 \nonumber \\
& & + Q^{-6} [ q^6 \xi^4 - 2 q^4 \xi^4 - q^2 \eta^4  
+ 10  q^2 \xi^2 \eta^2 ] \\
\Delta {\cal T}_{y^2} &=& {- 1 + q^2\over q^4} -10 {\eta^2 \over q^2}
\nonumber \\
& & + Q^{-6} 
\left[ q^4 \xi^4 - \eta^4 + 2 q^2 \eta^4  - 10 q^4 \xi^2 \eta^2 \right].
\end{eqnarray}

\section{The Inclusion of an On-Axis Shear}

\label{sec:OnAxisShear}

We are able to transform the entire results we have obtained so far
if the shear is aligned with the symmetry axis of the lens potential, i.e., an on-axis shear $\gamma$. The lens equation in this case is given by
\begin{eqnarray}
\xi &=&  x + \gamma x - a x  (x^2+y^2/q^2)^{\beta/2-1}  \label{eq:xi+shear} \\
\eta &=& y - \gamma y - {a y \over  q^2} (x^2+y^2/q^2)^{\beta/2-1}
\label{eq:eta+shear}
\end{eqnarray}
We now show that these equations can be transformed into eqs.  (\ref{eq:xi}) and (\ref{eq:eta}) by a simple transformation.
If we denote the transformed quantities with a prime it yields
\begin{eqnarray}
\xi^\prime &=& { \xi \over \sqrt{1 + \gamma} }  \qquad 
\eta^\prime = { \eta \over \sqrt{1 - \gamma} } \\
x^\prime &=& x \sqrt{1+\gamma} \qquad y^\prime = y \sqrt{1-\gamma} \\
a^\prime &=& a (1+\gamma)^{-\beta /2}  \qquad q^\prime = q \sqrt{ 1-\gamma
  \over 1 +\gamma }   
\end{eqnarray} 
whereas $\beta^\prime = \beta$ remains unchanged.
Applying the above transformation, we obtain 
\begin{eqnarray}
\xi^\prime &=&  x^\prime - a^\prime x^\prime  
(x^{\prime 2}+y^{\prime 2}/q^{\prime 2})^{\beta^\prime /2-1}   \\
\eta^\prime &=& y^\prime - {a^\prime y^\prime \over  q^{\prime 2}} 
(x^{\prime 2}+y^{\prime 2}/q^{\prime 2})^{\beta^\prime /2-1},
\end{eqnarray}
identical to equations (\ref{eq:xi}) and (\ref{eq:eta}) if we drop the prime.
We note that the magnification transforms like
\begin{equation}
\mu_i^\prime = \mu_i (1-\gamma^2)
\end{equation}
and the time delay transforms like
\begin{equation}
\tau^\prime = \tau - { \gamma \xi^2 \over (1+\gamma) } 
+ { \gamma \eta^2 \over (1-\gamma) }
\end{equation}
where we have used eq. (\ref{eq:tau}).
Since we are only interested in the relative time delay between
two images the additional terms in the transformation cancel out
and we obtain
\begin{equation}
\Delta \tau_{i,j}^\prime = \Delta \tau_{i,j}
\end{equation}
This means that if we assume an elliptical power-law potential the
time delays between the images remain unchanged under
the above transformation. Using only the time delays
we would not be able to disentangle the contribution
of the on-axis shear $\gamma$ and axis ratio $q$.

\section{The rotation angle of the lensing galaxy}
 \label{sec:rotation}
 
Often the positions of the images and the lensing galaxy
are given relative to one image of the quadruple lens or the lensing galaxy. Also the usual RA and DEC coordinate systems may not be aligned with the symmetry axes of the lensing galaxy, 
and so that we can not apply the derived formulae 
in \S\ref{sec:moments} directly. Since we assume an elliptical potential
we are able to derive the rotation angle of the major axis due
to eq. (7) of \citet{Wi96}.

Let $\Delta \alpha_i = a_i$ be the right ascensions and $\Delta \delta_i = d_i$
the declinations of the observed relative image positions relative to image 1
(image A) respectively. The rotation angle of the major axis is then given by
\begin{equation}
\tan( 2\theta) =  { 2( a_2 d_2 k_{34} + a_3 d_3 k_{42} 
+ a_4 d_4 k_{23} ) \over (a_2^2-d_2^2) k_{34} + (a_3^2 - d_3^2)
k_{42} + (a_4^2 - d_4^2) k_{23} }
\label{eq:C1}
\end{equation}
with $k_{ij} = a_i d_j - a_j d_i$ and $i,j = 2, 3, 4$ 
which corresponds to the images B, C and D. For RXJ\, 1131-1231 we obtain
from Table \ref{tab:RXJ}  $\tan(2\theta)= -0.5945$ which correspond to
a rotation angle of $\theta =-15.37^\circ$.

We note here that the rotation angle has an ambiguity of $90^\circ$
since $\theta$ is limited to $-45^\circ \leq \theta \leq 45^\circ$.
However, this ambiguity can be easily resolved as long as the
lensing galaxy can be observed (see \citealt{Wi96} for further discussion).
We finally arrive at the unrotated image positions by
\begin{equation}
\left( \matrix{ x_i \cr y_i } \right) = 
\left( \matrix{
\cos\theta & \sin\theta \cr 
-\sin\theta & \cos\theta } \right)   
\left( \matrix{ a_i - a_G \cr d_i - d_G } \right) 
\end{equation}
where $(a_G, d_G)$ is the relative galaxy position. 
Note that $\theta$ is measured relative to the $x$-axis. 
In observations the axis of the lensing galaxy is measured
north-to-east $\theta_{\rm N/E}$ which is given 
by  $\theta_{\rm N/E} = -90^\circ -\theta = -74.63^\circ$ for this case.
The rotation angle agrees quite well with the results of \citet{Cla06} (Table 4)
for their singular isothermal ellipsoid models with pure ellipticity and/or
on-axis shear.

\section{Estimation of the Slope of the Lens Potential}
\label{sec:slope}

When the external shear is negligible for an observed quadruple lens
we can derive an approximate estimate on the slope $\beta$ of the potential
using just the 4 image positions and the 3 time delays.
Following eq. (25) in \citet{WMK00}, we can write for the time delay
\begin{equation}
\Delta \tau_{i,j} = {2 \over B} (r_j^2 -r_i^2) + { (2-B) \over B} 
[\xi (x_i - x_j) + \eta ( y_i - y_j)],
\end{equation}
where we expressed $\beta$ in terms of $B$.
Since the ratios of the time delays are independent of the distance parameters
we can write 
$\Delta t_{1,2} / \Delta t_{1,3} = \Delta \tau_{1,2} / \Delta \tau_{1,3}$
for example.
Finally we use all the three time delay combinations to eliminate
$\xi$, $\eta$ or $B$. 
Now we obtain three equations in the form of
\begin{eqnarray}
c_1 \xi + c_2 \eta &=& 0 \nonumber \\
(B-2) c_3 \eta - 2 c_1 &=& 0 \nonumber \\
(B-2) c_3 \xi + 2 c_2 &=& 0 
\end{eqnarray}
where
\begin{eqnarray}
c_1 &=& \left| \matrix{
0 & \Delta t_{1,2} & \Delta t_{1,3} & \Delta t_{1,4} \cr 
x_1 & x_2 & x_3 & x_4 \cr
r_1^2 & r_2^2 & r_3^2 & r_4^2 \cr
1 & 1 & 1 & 1 \cr
} \right|,  \nonumber  \\
c_2 &=& \left| \matrix{
0 & \Delta t_{1,2} & \Delta t_{1,3} & \Delta t_{1,4} \cr 
y_1 & y_2 & y_3 & y_4 \cr
r_1^2 & r_2^2 & r_3^2 & r_4^2 \cr
1 & 1 & 1 & 1 \cr
} \right|, \nonumber  \\
c_3 &=& \left| \matrix{
0 & \Delta t_{1,2} & \Delta t_{1,3} & \Delta t_{1,4} \cr 
y_1 & y_2 & y_3 & y_4 \cr
x_1 & x_2 & x_3 & x_4 \cr
1 & 1 & 1 & 1 \cr
} \right|  
\end{eqnarray}
are the determinant of different combinations of the time delays
and the image positions.
Further we recall that for a pure elliptical potential
the relation
\begin{equation}
\xi \eta (x_i y_j - x_j y_i) + x_j y_j (\xi y_i - x_i \eta ) 
+ x_i y_i( x_j \eta - \xi y_j) = 0
\end{equation}
holds for $i,j=1,2,3,4$ and $i\neq j$ (\citealt{Wi96}).
This enables us to eliminate $\xi$ and $\eta$ successively.
Since we need here in principle only 2 image positions
we may actually obtain for a complicated system different
values of $B$ from different combinations of image
positions. We average over all 4 image positions to obtain an estimate of $\beta$
\begin{equation} \label{eq:B}
B = 2 + 2 {c_1 c_2 e_3 \over c_3 ( c_1 e_1 + c_2  e_2 ) }
\end{equation}
where
\begin{eqnarray}
e_1 &=& \left| \matrix{
-1 & 1 & -1 & 1 \cr 
x_1 & x_2 & x_3 & x_4 \cr
x_1 y_1 & x_2 y_2 & x_3 y_3 & x_4 y_4 \cr
1 & 1 & 1 & 1 \cr
} \right|, \nonumber  \\
e_2 &=& \left| \matrix{
-1 & 1 & -1 & 1 \cr 
y_1 & y_2 & y_3 & y_4 \cr
x_1 y_1 & x_2 y_2 & x_3 y_3 & x_4 y_4 \cr
1 & 1 & 1 & 1 \cr
} \right|,  \nonumber \\
e_3 &=& \left| \matrix{
-1 & 1 & -1 & 1 \cr 
y_1 & y_2 & y_3 & y_4 \cr
x_1 & x_2 & x_3 & x_4 \cr
1 & 1 & 1 & 1 \cr
} \right|.
\end{eqnarray}
The advantage of eq. (\ref{eq:B}) is that it is completely scale free.
We can immediately substitute the image positions and time delays without 
worrying about the units. However, we need to rotate the coordinate system
by the angle of the lens galaxy $\theta$ (see eq. \ref{eq:C1}), which points to 
a disadvantage of eq. (\ref{eq:B}) since
the uncertainty on $\theta$ enters the slope of the potential directly.

For RX J1131$-$1231 we obtain for a rotation angle of $\theta_G = -15.37^\circ$, 
$B=3.59$ which correspond to $\beta = 1.28$.
This prediction seems to overestimate the slope of the potential
because $\Delta t_{AD}$ is reduced to about 
$\Delta t_{AD} \approx 80$
days. We can adjust the value of $B$ so that the predicted time delays better match
the observed values. We find that a value of $B=2.65$ ($\beta = 1.14$) seems to be more appropriate
which predicts $\Delta t_{AD} \approx 93$ days comparable to the value in
Table \ref{tab:RXJ}. 
However, the large predicted offset is due to the neglect
of an external shear. It would be desirable to extend
eq.(\ref{eq:B}) in such a way that it would include an off-axis shear.

It is interesting that the system seems to be close to isothermal, 
but not exactly. This explains why we cannot account for the 
observed time delays very well. The conclusion is consistent with
the more advanced model in \citet{Suyu13} where they
also had to allow a small deviation
of the slope of the potential from isothermal (which corresponds to roughly $\beta \approx 1.05$ in our notation), but also include an external
shear in their model.

\label{lastpage}
\end{document}